# Wafer-level substrate-free low-stress silicon nitride platform for THz metadevices and monolithically integrated narrowband metamaterial absorbers

*Zhigang Li, Jiarui Jia, Wenjing Jiang, Wen Ou, Bo Wang, Xubiao Peng and Qing Zhao\**

Zhigang Li, Jiarui Jia, Xubiao Peng, Qing Zhao

Center for Quantum Technology Research and Key Laboratory of Advanced Optoelectronic Quantum Architecture and Measurements (MOE), School of Physics, Beijing Institute of Technology, Beijing 100081, China
E-mail: qzhaoyuping@bit.edu.cn

Zhigang Li, Wenjing Jiang, Wen Ou

Institute of Microelectronics of The Chinese Academy of Sciences, Beijing, 100029, China

Bo Wang

Beijing National Laboratory for Condensed Matter Physics, Institute of Physics, Chinese Academy of Sciences, Beijing 100190, China

Qing Zhao

Beijing Academy of Quantum Information Sciences, Beijing, 100193, China

The implementation of terahertz (THz) wafer-level metadevices is critical to advance the science for applications including (I) integrated focal plane array which can image for biology and (II) integrated narrowband absorbers for high spectral resolution THz spectroscopy. Substantial progress has been made in the development of THz metamaterials; however, a wafer-level low-stress THz metadevices platform remains a challenge. This paper experimentally demonstrates a substrate-free THz metadevices platform adopting engineered Si-rich and low-stress silicon nitride ($SiN_x$) thin films, achieving an extensive THz transparency up to $f = 2.5$ THz. A new analytical model is first reported from the Lorentz model that can accurately predict spectral responses of metal insulator metal (MIM) metamaterial absorbers. The model is experimentally validated in the THz range and exploited for the first demonstration of a THz absorber, which exhibits performance approaching the predicted results. Our results show that the wafer-level $SiN_x$ platform will accelerate the development of large-scale, sophisticated substrate-free THz metadevices. The Lorentz model and its quadratic model will be a very practical method for designing THz metadevices.

1. Introduction

$SiN_x$ films are known for their excellent performance in microelectromechanical system (MEMS) and in microphotonics due to their advantages of extremely low optical loss, low-stress, high optical nonlinearity, mechanical robustness, and strong chemical stability. Moreover, $SiN_x$ is demonstrated to be fully compatible with very large-scale integration processes, and is capable of being deposited on a variety of substrates. [1-3] $SiN_x$ films are usually deposited by low-pressure chemical vapor deposition (LPCVD) or plasma-enhanced chemical vapor deposition (PECVD) process. In the LPCVD process, it is known that films ranging from the stoichiometric composition with high tensile residual stresses to silicon-rich compositions with lower levels of tensile stress can be obtained by varying the source gas flow rates in the reaction chamber. Both types of films are amorphous.[4-5] The PECVD process is widely used owing to its high deposition rate at low process temperatures with excellent film uniformity.[3, 6] All these features make $SiN_x$ a versatile potential candidate for terahertz (THz) advanced optical device platforms. Some optical devices, such as metamaterials,[7-8] metadevices,[9-10] metasurfaces,[11-12] and microphotonics, [6, 13] have shown revolutionary growth in past decades, and the semiconductor materials and micro/nanofabrication technology are one critical factor that promotes their sucess.[14-16] However, there is an urgent demand to move the current proof-of-concept stage toward a large-scale for practical applications. Therefor to achieve large-scale micro/nanofabrication and high performance, wafer-level process, low-stress film, excellent optical ability, and substrate-free structure are required for these advanced optical devices. Especially the low-stress and substrate-free $SiN_x$ is widely used in reconfigurable metamaterials,[17-19] reconfigurable metasurface,[1, 12, 20] nanopore,[21-24] TEM sample stage,[5, 25-26] and sensors,[27-28] among others.

To meet this demand of optical devices and their applications, the chemical composition ratio between Si and N with the LPCVD process is engineered to reduce its mechanical stress and improve the ability of alkali corrosion resistance. In this research, an optimized 4-inch wafer-level substrate-free $SiN_x$ film with a tensile stress of as low as 40 MPa was investigated with the LPCVD process using a gas mixture of silane ($SiH_4$), ammonia ($NH_3$), and nitrogen ($N_2$). Finally, the $SiN_x$ was integrated with Au/$SiN_x$/Au metal-insulate-metal (MIM) metamaterial absorbers demonstrating a typically high absorption of 99% at 1.13 THz. Energy dispersive spectroscopy (EDS) and atomic force microscope (AFM) are utilized in both surface mapping and depth scanning modes to evaluate the compositional uniformity. THz time domain spectroscopy (THz-TDS) was utilized to investigate the THz properties of up to 2.5T. Experimentally fabricated narrowband metamaterial absorbers were compared to numerical simulations and analytical models. Remarkably narrow band and substrate-free film illustrated that the wafer-level low-stress $SiN_x$ platform was suitable for THz metadevices.

2. Results and discussion

Among the various $SiN_x$ films, silicon-rich films were selected because they possess superior properties such as low mechanical stress, high alkali corrosion resistance, and uniform film composition, meeting the requirements for THz metadevices. The $SiN_x$ film on a 4-inch silicon substrate is prepared by the LPCVD process. The Experimental Section discusses the detailed recipes for the $SiN_x$. The chemical composition of the SiNx film is characterized by EDS, as shown in **Figure** 1(a). An atomic ratio of Si:N of 58:42 is obtained, indicating that the low-stress (typically 40 MPa) $SiN_x$ is a silicon-rich $SiN_x$ film. For large-scale applications, the film is investigated through the measurement of surface roughness and potassium hydroxide (KOH) etching. Noncontact AFM was used to measure surface roughness. As shown in **Figure** 1(b), the surface roughness, Sa, of the $SiN_x$ film is 1.8 nm. The substrate is removed by KOH etching. After KOH etching, the surface roughness of the $SiN_x$ film becomes rougher, and as shown in **Figure** 1(c), the value of Sa becomes 2.7 nm. As shown in **Figure** 1(d), the 4-inch wafer-level substrate-free $SiN_x$ film is transparent. The membrane is entirely flat because its stress is as low as 40 MPa which was measured by the stress measurement system (KLA Tencor FLX 2320s) based on the Stoney equation.

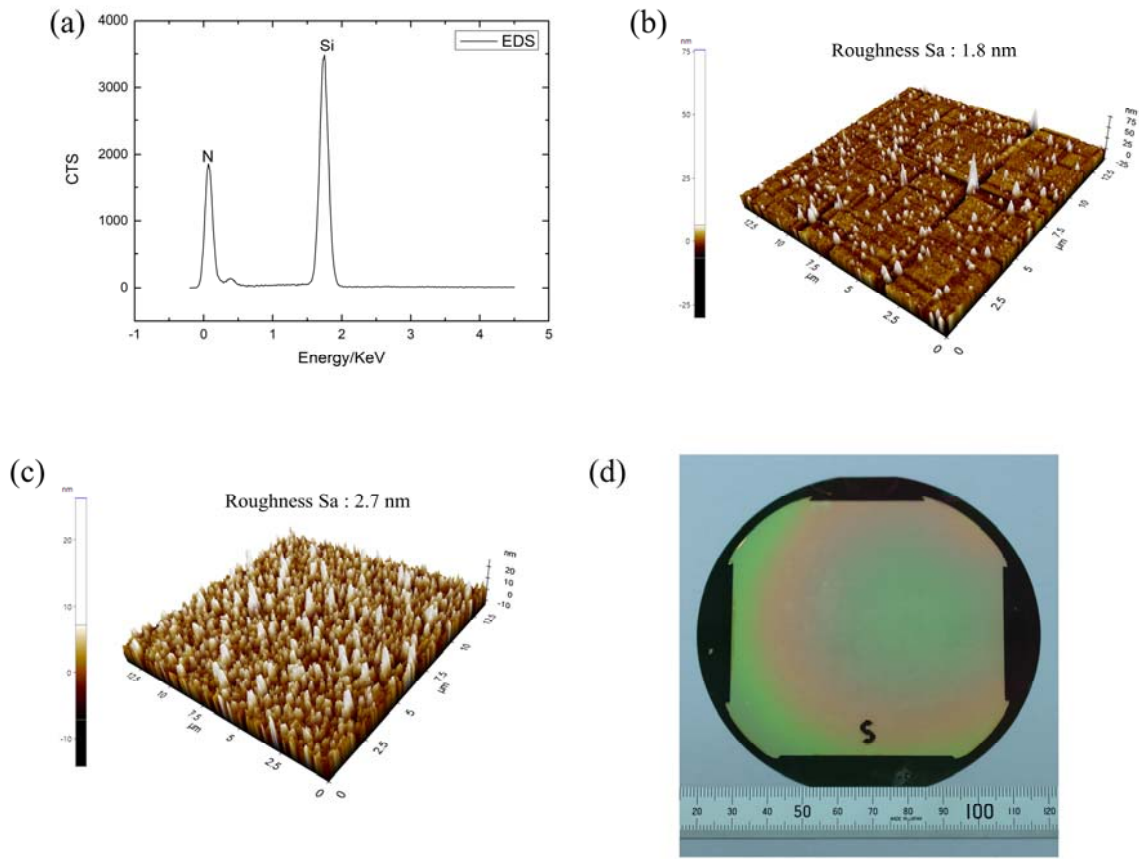

**Figure** 1. Wafer-level substrate-free low-stress SiN$_x$ film. (a) EDS analyses show the atomic ratio of Si:N of 58:42. (b) Surface roughness Sa was measured using noncontact AFM with a 13 × 13 μm of scan area. (c) Surface roughness Sa of scan area after KOH etching. (d) The 4-inch wafer-level SiN$_x$ film and its substrate are removed by KOH etching.

The THz properties, including both the index of refraction **n** and transmission **τ** of the SiN$_x$ film were characterized using THz-TDS, a spectroscopic technique that measures and analyzes the complete dielectric function of various materials in the THz region. **Figure** 2 shows the THz properties of the SiN$_x$ film. **Figure** 2(a) and (b) shows the time and frequency domain of the THz pulse transmitted through SiN$_x$ film, respectively. **Figure** 2 (c) shows the **n** and **τ** plots of the SiN$_x$ film. A comprehensive characterization is accomplished, showing that **n** ≈ 2.5 in 0.7~2.5THz and **τ** ≈ 0.99 in 0.2 ~ 2.5THz. The almost constant **n** and **τ** over a broad THz spectral range offer the advantage of low dispersion and low propagation loss that are desired for many THz metamaterials and metadevices. Thus, it was demonstrated in this study that the SiN$_x$ film reveals high transparency from the 0.2 to the 2.5 THz region. With the benefit of low stress, high alkali resistance, low-dispersion, and propagation loss, the wafer-level $SiN_x$ film provides a unique platform for large-scale fabrication of THz metadevices.

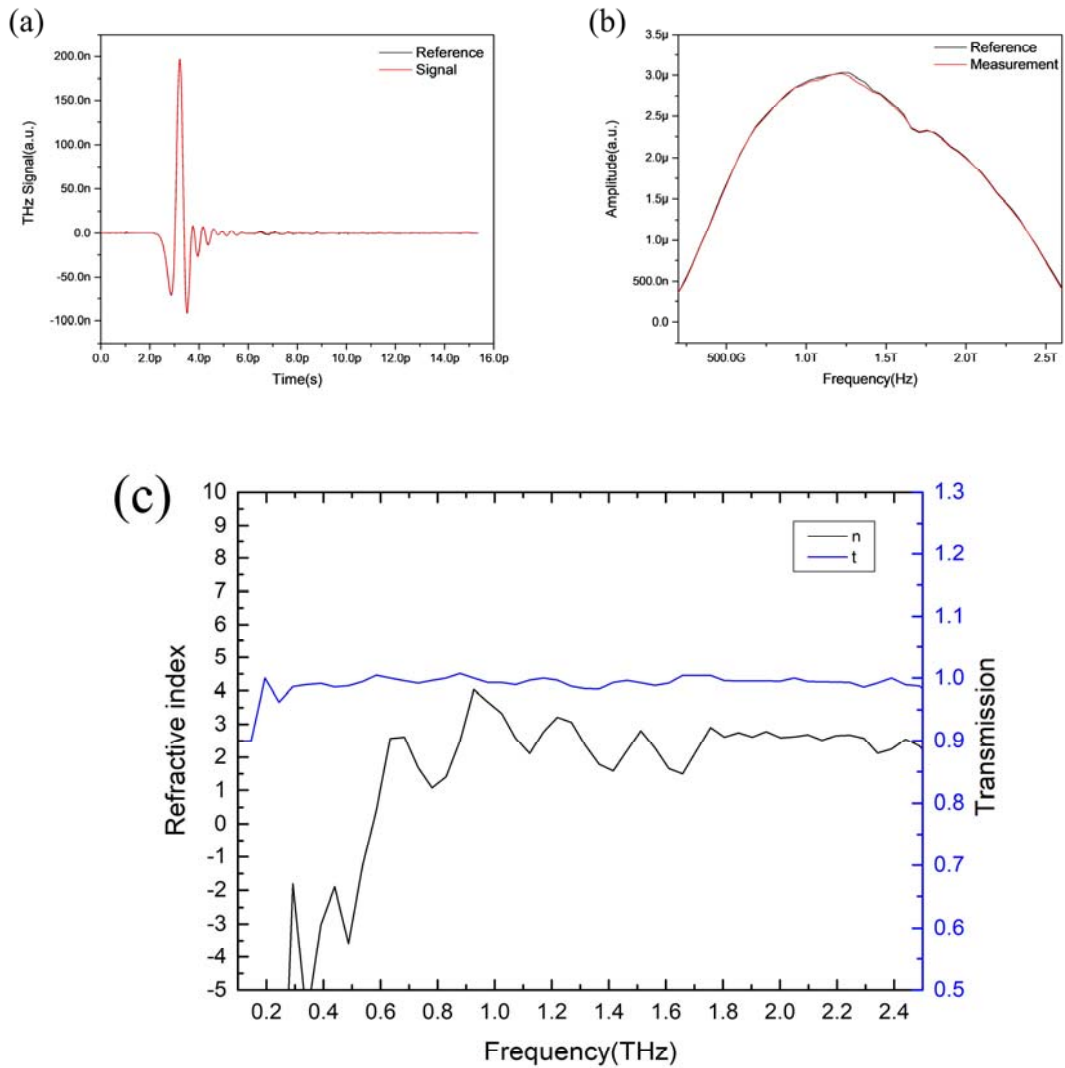

**Figure** 2 THz properties of the $SiN_x$ film. Time (a) and frequency (b) domains of a THz pulse transmitted through the $SiN_x$ film. (c) Refractive index and transmission of the $SiN_x$ film.

THz narrowband metamaterial absorbers are set up to evaluate the performance of the $SiN_x$ film. The proposed absorbers are composed of a three-layer metal-insulator-metal (MIM) metamaterial, an array of gold (Au), and a continuous ground layer separated by a subwavelength thickness $SiN_x$ layer, as shown in **Figure** 3(a). The top structures are square patches, with the lateral dimensions denoted as L (patch sides), and T (periodicity). The thickness of $SiN_x$ (t = 2μm) and Au ($t_{Au}$ = 100nm) was appropriately chosen for strong surface plasmon confinement and a wide lithographical tunability of the spectral absorption band in THz (0.5~2.5THz). To design a low-stress, substrate-free, and narrowband metamaterial absorber, absorption simulations are performed using finite integrate method calculations. When the THz waves are incident on the absorber, the electric dipole resonance is excited in the Au square patch array, inducing antiparallel currents on the Au ground layer via plasmonic coupling in the subwavelength dielectric gap, as shown in **Figure** 3(b). The magnetic field intensity, indicates that the current loop gives rise to strong magnetic field confinement. Moreover, near-unity absorption is achieved at the THz spectral band with the right level of loss to support critical coupling and conjugate matching. As shown in **Figure** 3(c), the absorption characteristic of the absorber was simulated, and fitted well with the Lorentz line shape function (Equation(1)). This function is usually used to characterize the line shape of infrared absorption spectra. [29-31] However, the authors have not found any reports regarding the Lorentz lineshape of THz absorption spectra. The evaluation of the best-fitting model is based on the values of three parameters, namely, center frequency ($f_c$), full width at half maximum ($w$), and peak absorptance ($H$). $y_0 = 0$ is the initial at $f = 0$. The coefficient of determination $R^2$ is closest to 1. **Figure** 3(d) summarizes all results.

$$\alpha(f) = y_0 + (H - y_0) \times \frac{w^2}{4(f - f_c)^2 + w^2} \tag{1}$$

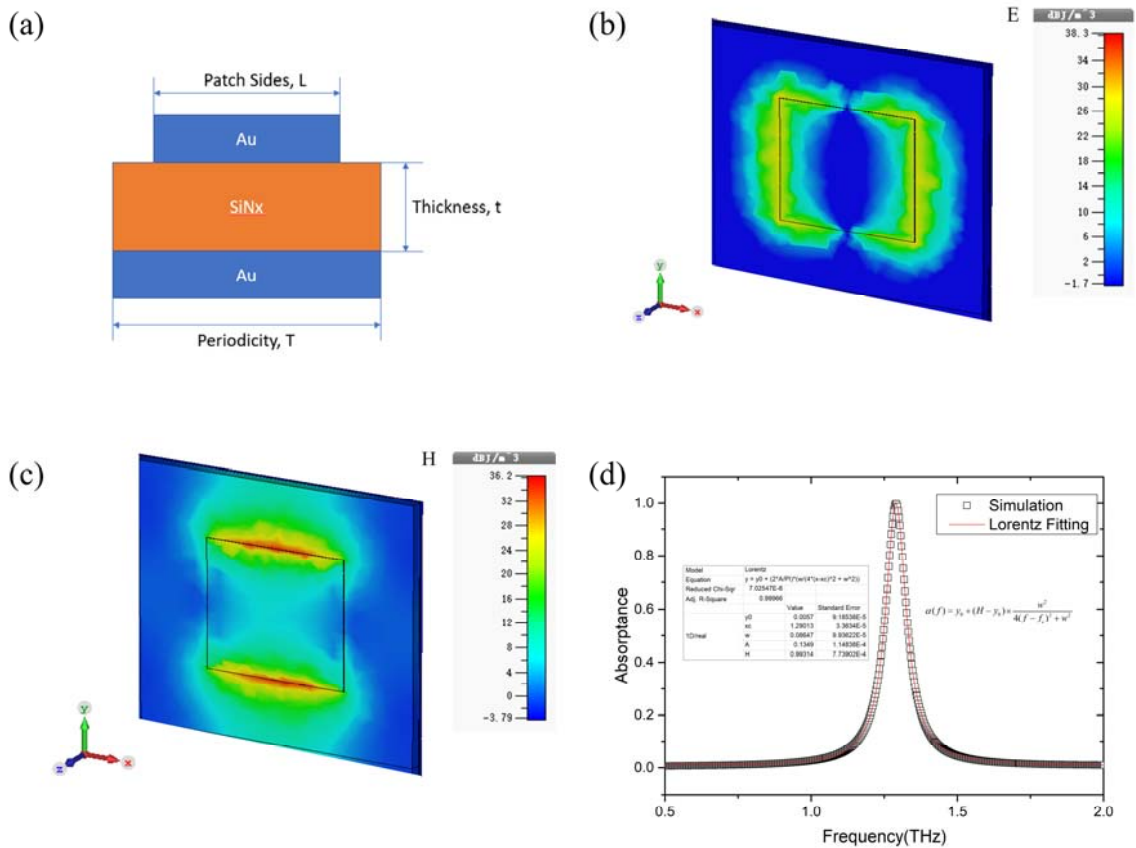

**Figure** 3 Narrow bandwidth MIM THz absorber. (a) Structure of the absorber unit used in the calculations. (b) Simulated electric dipole resonance. (c) Simulated magnetic field intensity. (d) Simulated absorptance curve with the Lorentz fitting.

To achieve narrow spectral selectivity and near-unity absorption, the center frequency, peak absorptance, and full width at half maximum (FWHM) of the fundamental resonance mode were analyzed with respect to the geometric parameters, as shown in **Figure** 4. It is worth noting that different patch sides, dielectric thickness and periodicity are required for optimized performance beyond this spectral range. The Lorentz model serves as a powerful tool to analyze the spectral responses and design narrowband metamaterial absorbers. This model comprises the absorption curve parameters as functions of patch sides, periodicity, and dielectric thickness. Hence, it provides a helpful understanding of the THz absorption condition with respect to the physical parameters of the absorbers. Consequently, the Lorentz model parameters ($f_c$, $w$, $H$) have been extracted by fitting the simulation data. The confidence level of each parameter is over 95%. This quadratic model analysis method is often used in forecasting[32-33]. Thus, the Lorentz quadratic model can be used to accurately predict the frequency at which maximum absorption occurs, the actual absorptance, and the bandwidth. The values of the absorption curve parameters, which are determined using the simulation data,

$$f_c(L) = 0.60 + 5.04 \times e^{\frac{-L}{20.1}} \tag{2}$$

$$w(L) = -0.66 + 0.79 \times e^{\frac{-L}{717.2}} \tag{3}$$

$$H(T) = 6.36^{-9} \times (T + 55.2)^{5.6} \times e^{\frac{-90.5-T}{19.2}} \tag{4}$$

$$w(T) = 0.04 + 0.226 \times e^{\frac{-T}{31.1}} \tag{5}$$

$$H(t) = -1.75 + 2.76 \times e^{\frac{-(t-1.97)^2}{6.99}} \tag{6}$$

$$w(t) = 0.09 + 0.454 \times e^{\frac{-t}{0.28}} \tag{7}$$

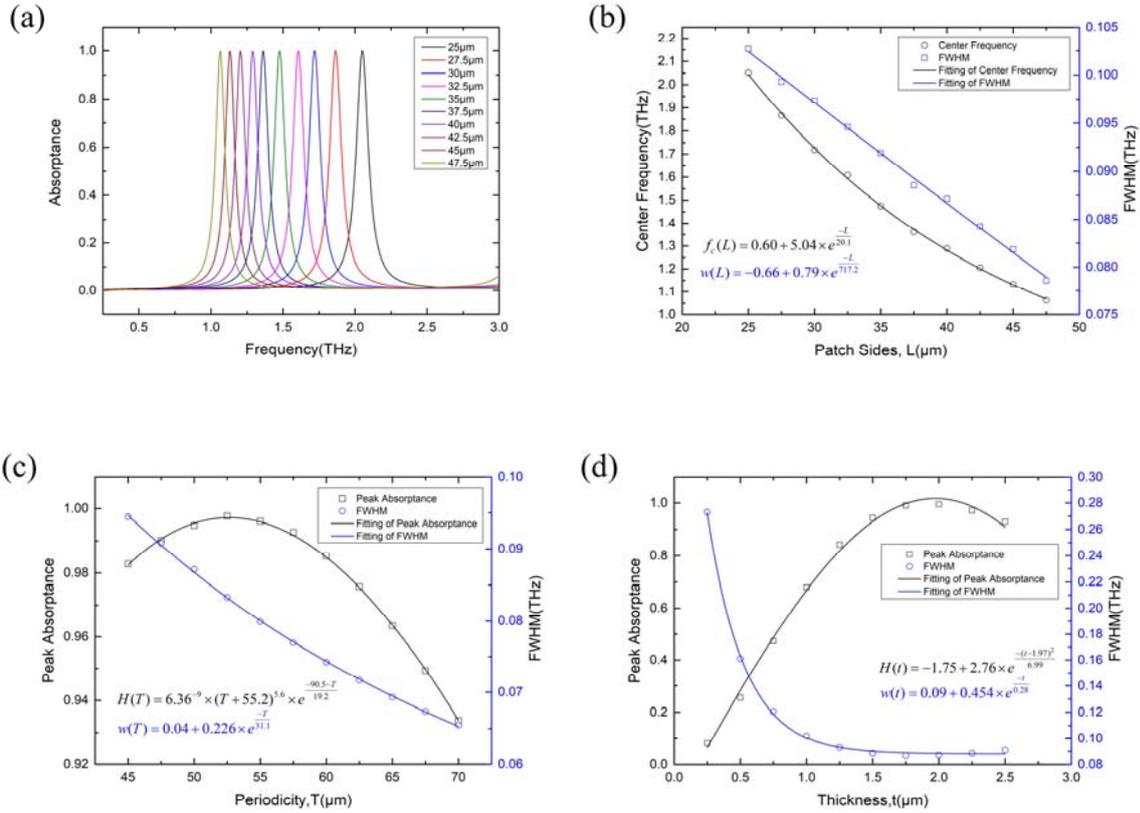

**Figure** 4 Optimization of MIM THz absorbers with respect to the geometrics. (a) Simulated the THz absorption responses of the lithographically tuned MIM metamaterial absorbers with optimized geometric dimensions. (b) Simulated center frequency and FWHM with respect to the patch sides. (c) Simulated peak absorptance and FWHM with respect to periodicity. (d) Simulated peak absorptance and FWHM with respect to thickness.

The above expressions for the absorption curve elements are expressed in terms of geometric variables (L, T, t), which can guide the design and optimization of the metamaterial absorbers. Equations (4, 6) indicate that the highest absorption can be achieved by optimizing the variables(T, t).

As shown in **Figure** 5, the MIM THz absorbers were fabricated and characterized on the same substrate, with lateral dimensions (L, T, t) as the only variables. The geometric parameters of each absorber are optimized to achieve high selectivity based on the model proposed in this study. Since the center frequency ($f_c$) is inversely proportional to the patch side (L) and almost independent of other dimensions (Equation (2)), the center frequency can be lithographically tuned by choosing an appropriate L while consistently maintaining a narrowband absorption. It is worth noting that all fabricated devices consistently demonstrate high absorptance (> 90%) and narrow bandwidth (FWHM < 0.25THz) over a wide spectral region, as shown in **Figure** 5(a). The top plasmonic square patch structures were patterned via a liftoff process, followed by sputtering of 10/100 nm Ti/Au film. **Figure** 5(b) shows the scanning electron microscopy (SEM) image of the MIM THz absorber. The inset is a cross-sectional SEM image. As seen from the image, only the films Au/SiN$_x$/Au/SiN$_x$ stacks are shown clearly, because the silicon substrate is removed. **Figure** 5(c) shows the comparison between the Lorentz model and the measurement results. The inset image highlights that an absorption bandwidth of only 0.11 THz and the near-perfect absorption (> 99%) are simultaneously obtained, while strong reflectance (≈ 0%) is achieved in the out-of-band spectral region. As shown in **Figure** 5(d), the measured center frequency is almost consistent with the Lorentz quadratic model prediction. The measured FWHM is a little larger than the prediction. It may be probable that the incident angle affects the FWHM, because the THz incident angle is fixed at 30° due to the mechanical limitation of the THz-TDS. And the asymmetric lineshape of the THz absorption spectra with the incident angles was reported in a previous research.[34] It is also worth noting that Lorentz and its quadratic model accurately predict the spectral response of the absorber from which one can obtain the bandwidth, center frequency, and absorptance levels.

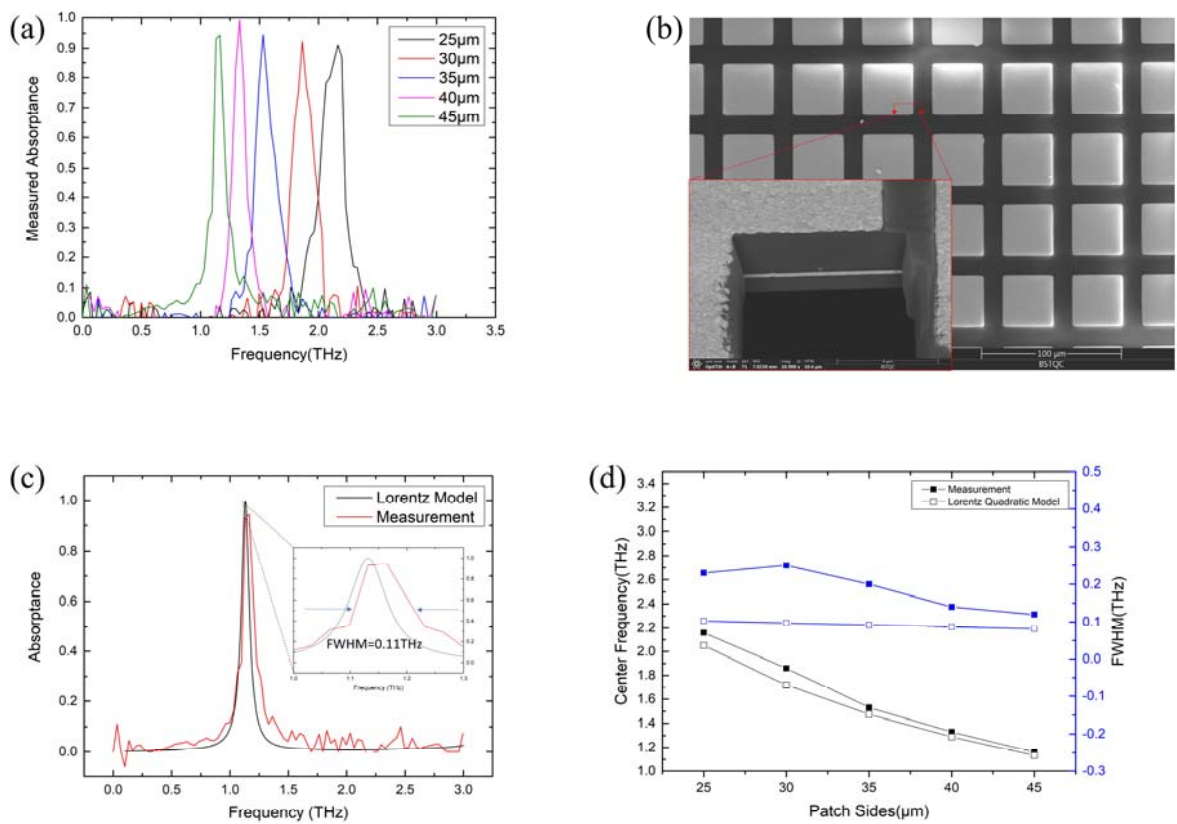

**Figure** 5 Experimental demonstration of narrowband MIM THz absorbers. (a) Absorption spectra of the five narrowband MIM THz absorbers with lithographically tuned peak absorption. (b) The SEM image of the narrowband absorbers. The inset is a cross-sectional SEM image. (c) Comparison of the absorption spectrum

predicted by the Lorentz model (black line). The inset highlights the narrow FWHM that reaches its theoretical limit of 10%. (d) Experimental verification of center frequency and FWHM of the five absorbers predicted using the Lorentz quadratic model.

3. Conclusion

This paper demonstrates wafer-level substrate-free THz metadevices platform using specially engineered $SiN_x$ films. The prepared $SiN_x$ film has a wide THz transparency, which covers $f$ = 0.2 up to $f$ = 2.5 THz measured from THz-TDS. Due to its low mechanical stress and high alkali-corrosion resistance, the $SiN_x$ film can reach wafer-level substrate-free film. Using the developed $SiN_x$ film, narrowband metamaterial absorbers with a new design model, i.e., the Lorentz model, were demonstrated in this study. Thus, a new monolithic THz wafer-level platform is provided, which will accelerate the development of THz metadevices.

4. Experimental Section

*Preparation of low-stress $SiN_x$ film*: The wafer-level low-stress $SiN_x$ films on a 4-inch silicon substrate were prepared by a Bruce LPCVD. The silicon source is dichlorosilane (DCS), and the nitrogen source is ammonia ($NH_3$). The deposition temperature is 900°C, and the total deposition pressure is set at 300 mTorr. A high DCS: $NH_3$ gas ratio of 2:1 is chosen to produce low-stress and high alkali resistance.[5]

*Device Fabrication Process*: Conventional MEMS processes are utilized to fabricate the THz narrowband absorbers on low-stress $SiN_x$ film. Firstly, a 5000-Å thick LPCVD $SiN_x$ layer is generated on double sides of the wafer. A 1000-Å gold (Au) layer is evaporated on the front side, and 2.0 μm of the PECVD $SiN_x$ layer is deposited on top of the Au layer. Next, front lithography is performed to pattern the absorption layer. The Au layer is evaporated on the photoresists and is then, lifted off. Finally, lithography and reactive ion etching are performed to open the back etching window. The backside silicon substrate is released by KOH etching.

*THz Measurement System*: THz-TDS was used to characterize the THz properties of the $SiN_x$ film. The THz wave was generated using a low-temperature grown GaAs photoconductive antenna integrated with a Si hemisphere lens. The frequency is in the range of 0.2~2.5 THz, as measured using THz-TDS, with a central laser wavelength of 800 nm, the pulse duration of 100 fs, average power of 30 mW, and SNR of >65 dB.[35] THz detection was realized using a ZnTe crystal via electro-optical sampling. Both the transmitted and reflected THz waves were measured, and the absorption then equates to 1-R-T, where reflectivity and transmissivity are denoted as by R and T.

**Acknowledgments**

This work was supported by the National Key R&D Program of China (No. 2018YFC2000901 & No.2018YFC2000902).

Received: ((will be filled in by the editorial staff))
Revised: ((will be filled in by the editorial staff))
Published online: ((will be filled in by the editorial staff))